\documentclass[fleqn,10pt]{wlscirep}
\usepackage{amsmath,amssymb}
\usepackage{tensor}
\usepackage{array}
\usepackage{bm}

\title{Vibronic Boson Sampling: Generalized Gaussian Boson Sampling for Molecular Vibronic Spectra at Finite Temperature}

\author[1,*]{Joonsuk Huh}
\author[2,$\dagger$]{Man-Hong Yung}
\affil[1]{Department of Chemistry, Sungkyunkwan University, Suwon 440-746, Korea}
\affil[2]{Department of Physics, South University of Science and Technology of China, Shenzhen 518055, China}

\affil[*]{joonsukhuh@skku.edu}

\affil[$\dagger$]{yung@sustc.edu.cn}

\def\noteC#1{\textbf{\color{green}}} 

\newcommand{\diag}{\mathrm{diag}}

\newcommand{\tr}{\operatorname{Tr}}

\def\bra#1{\mathinner{\langle{#1}|}}
\def\ket#1{\mathinner{|{#1}\rangle}}

\begin{abstract}
Molecular vibroic spectroscopy, where the transitions involve non-trivial Bosonic
correlation due to the Duschinsky Rotation, is strongly believed to be in a similar 
complexity class as Boson Sampling. At finite temperature, the problem is represented 
as a Boson Sampling experiment with correlated Gaussian input states. 
This molecular problem with temperature effect is intimately related to the various versions of Boson Sampling sharing the similar computational complexity. Here we provide a full
description to this relation in the context of Gaussian Boson Sampling.
We find a hierarchical structure, which 
illustrates the relationship among various Boson Sampling schemes.
Specifically, we show that every instance of Gaussian Boson Sampling
with an initial correlation can be simulated by an instance of
Gaussian Boson Sampling without initial correlation, with only a
polynomial overhead. Since every Gaussian state is associated with a
thermal state, our result implies that every sampling problem in
molecular vibronic transitions, at any temperature, can be simulated
by Gaussian Boson Sampling associated with a product of vacuum modes. 
We refer such a generalized Gaussian Boson Sampling motivated by the molecular sampling problem as \emph{Vibronic Boson Sampling}.
\end{abstract}
\begin{document}

\flushbottom
\maketitle
%
%
\thispagestyle{empty}


\section*{Introduction}
A quantum simulation protocol called ``Boson Sampling", proposed by Aaronson and Arkhipov~\cite{Aaronson2011}, represents a serious challenge to the extended Church-Turing thesis.
In Boson Sampling, indistinguishable and non-interacting photons are injected into a linear optical network; it results in a photon distribution, which is not classically accessible in sufficiently large scale.  

Compared with a fully-working quantum computer, Boson Sampling consumes less physical resources for implementation; it requires only passive optical devices and photon detectors. Recently, successful small-scale experimental realizations of Boson Sampling have been reported~\cite{Spring2013,Broome2013,Crespi2013,Tillmann2013}. However, the current technique is still far from demonstrating the quantum supremacy with large-scale implementations of Boson Sampling. For example, the preparation of the initial Fock states can only be achieved with an exponentially-small probability, without a reliable single-photon source. 

Several generalized Boson Sampling schemes, sharing similar computational complexity, have been reported. For example, the Scattershot Boson Sampling~\cite{Lund2014,Bentivegna2015} using two-mode squeezed states was suggested to overcome the difficulty in preparing the initial Fock states. Moreover, non-classical quantum optical states like displaced Fock states, photon-added or subtracted squeezed vacuum states, and cat states have been considered~\cite{Olson2015,Seshadreesan2015,Rohde2015}. In addition, a time-domain Boson Sampling method has been proposed~\cite{Motes2014,Pant2016} for a scalable implementation of Boson Sampling.

Apart from photons, non-optical Boson Sampling, including trapped-ion~\cite{Shen2014} and superconducting circuits~\cite{Peropadre2015} have been proposed theoretically, which aims to overcome the state-preparation problem, and can be readily extended to Gaussian input states~\cite{rahimi2015}, including squeezed vacuum and squeezed coherent states for molecular applications~\cite{huh2015}.

In particular, Rahimi-Keshari and coworkers~\cite{rahimi2015} considered a Gaussian extension of Boson Sampling, where the input states for the optical network are a product of single-mode Gaussian states e.g. thermal states (see also Ref.~\cite{PhysRevA.90.063836}) and squeezed vacuum states. It was found that, similar to the original version of Boson Sampling, simulating Gaussian Boson Sampling involving squeezed states is still a hard problem for classical devices. Moreover, the molecular problem proposed by Huh et al.~\cite{huh2015} represents an application of Gaussian Boson Sampling for sampling molecular transitions at zero temperature. 
Therein, quantum simulation of the molecular spectroscopy can be achieved by sending  uncorrelated single-mode squeezed coherent or squeezed vacuum states through an optical network. 
We note, here, recently, a small-scale trapped-ion simulation of the molecular problem has been successfully performed by Shen et al.~\cite{Shen2017}.

However, when simulating molecular transitions at finite temperatures, the input Boson modes to the linear optical network are in general correlated. Therefore, the molecular problems belong to a generalized Gaussian Boson Sampling problem, which involves the preparation of general Gaussian states, with mode correlation in general, as the input for Boson Sampling. To distinguish it from the Gaussian Boson Sampling studied in the literature~\cite{rahimi2015}, we refer to it as ``Vibronic Boson Sampling", as the study is motivated by the molecular sampling problem. Because Vibronic Boson Sampling captures the initial correlation not included in Gaussian Boson Sampling, the term may be used interchangeably with ``generalized Gaussian Boson Sampling".


Here we study the scenarios where initial correlation is included in Gaussian Boson Sampling and its relationship with other ways of performing Boson Sampling. In particular, we present a hierarchical framework that can reduce all problems in Boson Sampling~\cite{Aaronson2011}, Scattershot Boson Sampling~\cite{Lund2014,Bentivegna2015}, and Guassian Boson Sampling~\cite{rahimi2015} at any temperature, as an instance of Vibronic Boson Sampling at zero temperature, which is in turn equivalent to Gaussian Boson Sampling at zero temperature using the result in Ref.~\cite{huh2015}. In other words, we show that it is possible to absorb all the initial Gaussian correlation into the operations of Gaussian Boson Sampling without initial correlation. 
The cost is a doubling of the number of squeezed modes as Scattershot Boson Sampling; and it also requires doubling the photon network size, which is about four times~\cite{Reck94} more expensive than the zero Kelvin application.

On the other hand, since any Gaussian state can be prepared from a thermal state of Bosons, which normally requires sampling from the Boltzmann distribution of the Boson modes. In our approach, we can avoid sampling the initial thermal distribution by replacing the procedure with a squeezing operation acting jointly on the original modes and a set of ancilla modes in the vacuum state. Moreover, we show that such a squeezing operation can be incorporated into a multi-dimensional Bogoliubov transformation in the extended Hilbert space, the joint operation consists of sequential operations of displacement-rotation-squeezing-rotation operators~\cite{Braunstein2005,Cariolaro2016}.

\section*{Results}

\subsection*{Boson Sampling and the Gaussian version}
In general, the probability $P_{\hat{\Pi}}$ of a projective measurement $\hat{\Pi}$ for sampling Bosons, with a problem-dependent scattering operator $\hat{O}$ and an input state $\hat{\rho}_{\mathrm{in}}$, is given by, 
\begin{align}
P_{\hat{\Pi}}=\tr\left[\hat{O}\hat{\rho}_{\mathrm{in}}\hat{O}^{\dagger}\hat{\Pi}\right] . 
\end{align}
Specifically, when $\hat{\Pi}=\ket{\mathbf{m}}\bra{\mathbf{m}}$ is a photon-number projector where $\ket{\mathbf{m}}=\ket{m_{1},\ldots,m_{M}}$ (bold symbol labels column vectors), $\hat{\rho}_{\mathrm{in}}=\ket{\phi}\bra{\phi}$ is a pure state where $\ket{\phi}=\ket{1,\ldots,1,0,\ldots,0}$, and $\hat{O}=\hat{R}_{U}$ is the rotation operator of the unitary matrix $U$ generated by a linear optical network, the resulting problem represents the original Boson Sampling~\cite{Aaronson2011}. 

In Gaussian Boson Sampling~\cite{rahimi2015}, a product of Gaussian modes are employed as the input, which can be represented by the following squeezed thermal state, 
$\hat{\rho}_{\mathrm{in}}=\hat{S}_{\Sigma} \ \hat{\rho}_{\mathrm{th}} \ \hat{S}_{\Sigma}^{\dagger}$,
where ${\hat S_\Sigma }$ is a product of $M$ single-mode squeezing operators $\hat{S}_{\sigma_{k}} $, i.e.,
${\hat S_\Sigma } = \bigotimes_{k=1}^{M}{\hat S_{{\sigma _k}}}$,
and  $\Sigma=\mathrm{diag}(\sigma_{1},\ldots,\sigma_{M})$ is the (real-valued) squeezing parameter matrix, with ``diag" labeling a diagonal matrix. 
The thermal state $\hat{\rho}_{\mathrm{th}}$ is a product of individual thermal states with potentially different frequencies $\omega_i$ and temperatures ($\beta_{k}=1/k_{\mathbf{B}}T_{k}$), i.e., 
${{\hat \rho }_{{\text{th}}}} = {{\text{e}}^{ - \hat H}}/{\text{Tr}}({{\text{e}}^{ - \hat H}})$ with $\hat H = \sum\nolimits_{k = 1}^M {{\beta _k}} \hbar {\omega _k}\hat a_k^\dag {{\hat a}_k}$ defined by the Boson operators satisfying $[\hat{a}_{k},\hat{a}_{l}^{\dagger}]=\delta_{kl}$. 
The state is then sent through a linear optical network, with the action ${\hat R}_U$, which results in the following distribution,
\begin{equation}
P\left( {\mathbf{n}} \right) = {\text{tr}}[{{\hat R}_U}{{\hat \rho }_{{\text{in}}}}\hat R_U^\dag \left| {\mathbf{n}} \right\rangle \left\langle {\mathbf{n}} \right|] \ .
\end{equation}

\subsection*{Gaussian correlation}
In general, an arbitrary Gaussian state $\hat{\rho}_{\mathrm{G}}$ can be generated from a thermal states $\hat{\rho}_{\mathrm{th}}$ with a Gaussian operator $\hat{O}_{\mathrm{G}}$~\cite{Braunstein2005,Weedbrook2012,Adesso2014,Cariolaro2016}, i.e.   $\hat{\rho}_{\mathrm{G}}=\hat{O}_{\mathrm{G}}\hat{\rho}_{\mathrm{th}} \hat{O}_{\mathrm{G}}^{\dagger}$.  The input states considered in Scattershot Boson Sampling~\cite{Lund2014} and Gaussian Boson Sampling~\cite{rahimi2015} are special instances of this generalized multidimensional Gaussian state.  

Note that the action of a Gaussian operator $\hat{O}_{\mathrm{G}}$ on the Boson creation operator, $\hat{\mathbf{a}}^{'\dagger} = \hat{O}_{\mathrm{G}}^{\dagger}\hat{\mathbf{a}}^{\dagger} \hat{O}_{\mathrm{G}}$ can be generally defined~\cite{Weedbrook2012,Adesso2014} as follows, 
\begin{equation}
\hat{\mathbf{a}}^{'\dagger} = X\hat{\mathbf{a}}+Y\hat{\mathbf{a}}^{\dagger}+\mathbf{z} \ ,
\label{eq:genBT}
\end{equation}  
where the $M\times M$ matrices $X$ and $Y$ satisfy the following two conditions, $XX^{\dagger}-YY^{\dagger}=I$ and $XY^{\mathrm{t}}=YX^{\mathrm{t}}$. 
$\hat{\mathbf{a}}^{'\dagger}$ and $\hat{\mathbf{a}}^{\dagger}$ are the $M$-dimensional boson creation operator column vector of output and input states, respectively.                               
A shorthand notation~\cite{Ma1990} for the boson operator vector $\hat{\mathbf{x}}$ has been used, \emph{i.e.}
$\hat{A}\hat{\mathbf{x}}\hat{B}=(\hat{A}\hat{x}_{1}\hat{B},\ldots,\hat{A}\hat{x}_{M}\hat{B})^{\mathrm{t}}$.
The corresponding arbitrary unitary operator
 for the linear transformation in Eq.~\ref{eq:genBT} can be decomposed into sequential quantum optical operators that $\hat{O}_{\mathrm{G}}=\hat{D}_{\mathbf{z}^{*}}\hat{R}_{U_{\mathrm{L}}}\hat{S}_{\Sigma}\hat{R}_{U_{\mathrm{R}}^{\dagger}}$. Here $\hat{D}_{\mathbf{z}^{*}}$ is the displacement operator with the displacement vector $\mathbf{z}^{*}$. 
Accordingly, $X$ and $Y$ in Eq.~\ref{eq:genBT} are identified as $U_{\mathrm{L}}\sinh(\Sigma)U_{\mathrm{R}}^{\mathrm{t}}$ and $U_{\mathrm{L}}\cosh(\Sigma)U_{\mathrm{R}}^{\dagger}$, respectively~\cite{Braunstein2005,Cariolaro2016}, via the singular value decomposition (SVD). 
The detailed forms and the actions of the quantum optical operators can be found in Methods. 



In the following, to make a distinction from Gaussian Boson Sampling, we define \emph{Vibronic Boson Sampling} as the class of sampling problems utilizing the most general Gaussian states as the input. The name is motivated by the problem of sampling molecular vibrational transitions at finite temperatures (see Methods). The connection between Boson Sampling and molecular transitions was first made in  Huh et al.~\cite{huh2015}, where the mode correlations are absorbed into local operations. However, such a procedure is possible for Gaussian states associated with the vacuum state (i.e., zero temperature) only. 

\begin{figure}[htb]
\begin{center}
\includegraphics[width=0.45 \textwidth]{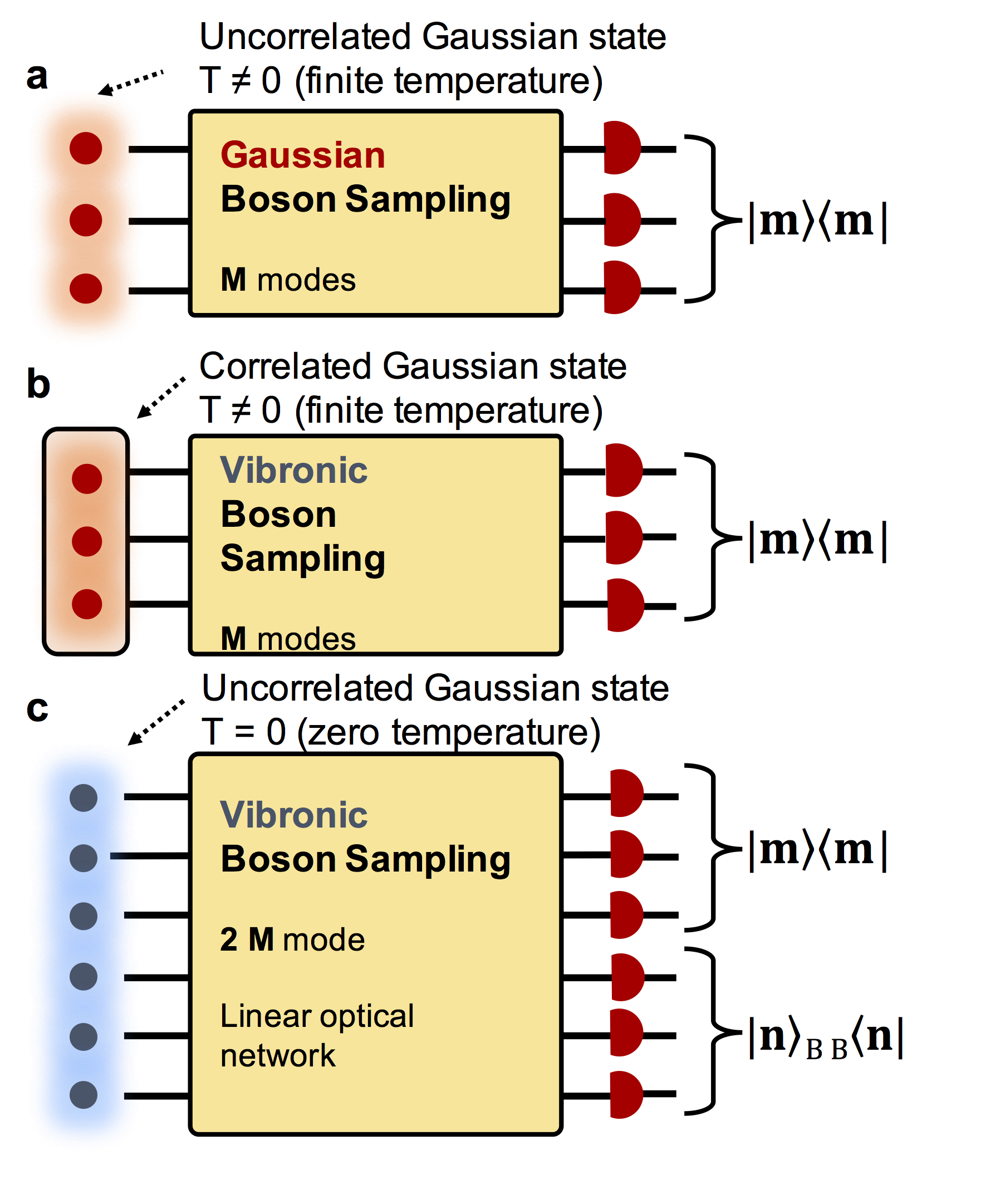}
\caption{Gaussian Boson Sampling device for thermal states. 
{\bf a}. The uncorrelated squeezed thermal state ($\hat{S}\rho_{\mathrm{th}}\hat{S}^{\dagger}$) enters into linear  photon network. The output Fock states ($\ket{\mathbf{m}}\bra{\mathbf{m}}$) are detected, and the input number states are not resolved. 
{\bf b}. The correlated squeezed thermal state ($\hat{S}\hat{R}\rho_{\mathrm{th}}\hat{R}^{\dagger}\hat{S}^{\dagger}$) enters into linear photon network. The output Fock states ($\ket{\mathbf{m}}\bra{\mathbf{m}}$) are detected, and the input number states are not resolved. 
{\bf c}. The thermal state is prepared as a pure state in the extended Hilbert space with the ancillary modes. The input states are prepared as uncorrelated squeezed coherent or vacuum input states. Both of output ($\ket{\mathbf{m}}\bra{\mathbf{m}}$) and input ($\ket{\mathbf{n}}_{\mathrm{B}}\mathrm{\tensor[_B]{\bra{\mathbf{n}}}{}}$) Fock states are resolved via measuring the photon output of the extended optical modes.  
}\label{fig:pictorial}
\end{center}
\end{figure}

\subsection*{Vibronic Transition and Franck-Condon Profile}
In the following, we are going to connect the problem of molecular spectroscopy with the vibronic extension of Boson Sampling. Such a connection was first pointed out in Ref.~\cite{huh2015}, but details are missing. For completeness, here we summarize a self-contained description and extend the result for initial thermal states.

Under the Born-Oppenheimer approximation, the total molecular wavefunction of the nuclear (${\bf \hat{R}} = (\hat{R}_1, \hat{R}_2, ...) $) and electronic (${\bf \hat{r}} = (\hat{r}_1,\hat{r}_2,...)$) degrees of freedom are separated, i.e., $  \psi \left( {{\mathbf{\hat{r}}},{\mathbf{\hat{R}}}} \right)\phi \left( {\mathbf{\hat{R}}} \right)$, and the electronic wavefunction $\psi \left( {{\mathbf{\hat{r}}},{\mathbf{\hat{R}}}} \right)$ depends parametrically the nuclear coordinates $\bf \hat{R}$. 

As a result, for transitions involving two electronic levels, $\ket{g}$ and $\ket{e}$, the molecular Hamiltonian $\cal{H}_{\rm mol}$ can be approximated as follows:
\begin{equation}
{\cal{H}_{{\text{mol}}}} = \left| g \right\rangle \left\langle g \right| \otimes {H_g} + \left| e \right\rangle \left\langle e \right| \otimes {H_e} \ ,
\end{equation}
where ${H_g} = \sum\nolimits_k \hbar{{\omega _k}\hat{a}_k^\dag {\hat{a}_k}} $ is the Hamiltonian of a set of phonon modes for the electronic ground state, and ${H_e} = \sum\nolimits_k \hbar{\omega'_k}\hat{a}'^{\dagger}_{k}  \hat{a}'_{k}  + \hbar{\omega _{{\text{ad}}}}$ is the Hamiltonian of the phonon modes for the excited electronic state. Here $\omega_{\rm ad}$ is called the electronic adiabatic transition frequency, and is usually set to be zero as an offset.   

In general, the two set of the boson operators are related by a unitary transformation, called Duschinsky transformation $\hat{U}_{\mathrm{D}}$~\cite{duschinsky:1937}, such that
\begin{equation}
\hat{a}'_{j} = {\hat{U}_{\mathrm{D}}^\dag}  \hat{a}_{j}  \hat{U}_{\mathrm{D}} = \sum\limits_j { ( {{X_{jk}}{\hat{a}_k} + {Y_{jk}}\hat{a}_k^\dag } )}  + {z_j} \ ,
\end{equation}
which reduces to the case of linear optics if all elements of $Y$ and $\bf z$ are zero.

When the molecule is driven by an external field at a given frequency $\omega_{\mathrm{ex}}=\omega_{\mathrm{v}}+\omega_{\mathrm{ad}}$, i.e., ${{\mathbf{E}}_0}\exp \left( {i\hbar\omega_{\mathrm{ex}} t} \right)$, the transition is determined by the matrix element: $\left\langle {e,{\bf m}'} \right|{{\mathbf{E}}_0}e{\mathbf{\hat{r}}}\left| {g,{\bf n}} \right\rangle  = {{\mathbf{E}}_0}e\left\langle {e,{\bf m}'} \right|{\mathbf{\hat{r}}}\left| {g,{\bf n}} \right\rangle$. Under the Condon approximation, which assumes the transition dipole moment to be independent of the nuclear motion, i.e., $\left\langle {e,{\bf m}'} \right|{\mathbf{\hat{r}}}\left| {g,{\bf n}} \right\rangle  = \left\langle e \right|{\mathbf{\hat{r}}}\left| g \right\rangle \left\langle {{\bf m}'} \right|\left. {\bf n} \right\rangle$, we can then focus on the matrix element involving only on the phonon modes,
\begin{equation}
\left\langle {{\bf m}'} \right|\left. {\bf n} \right\rangle  = \left\langle {\bf m} \right|\hat{U}_{\mathrm{D}}\left| {\bf n} \right\rangle \ ,
\end{equation}
which can be reduced to the problem of original Boson Sampling~\cite{Aaronson2011}, if we choose $\hat{U}_{\mathrm{D}}$ to contain a unitary rotation matrix encoded with a complex matrix, and start with the initial state $\left| {\bf n} \right\rangle  = \left| {111..100...0} \right\rangle $.

Finally, the molecular absorption lineshape is proportional to the so-called Franck-Condon profile (FCP), 
\begin{align}\label{FCP_thermal}
\mathrm{FCP}\left( \omega_{\mathrm{v}}  \right) = \sum\limits_{\mathbf{m}} {{{\left| {\left\langle {\mathbf{m}} \right|\hat{U}_{\mathrm{D}}\left| {\mathbf{n}} \right\rangle } \right|}^2}{P_{{\text{in}}}}\left( {\mathbf{n}} \right){\delta _{\omega_{\mathrm{v}}} (\Delta) }}  \ ,
\end{align}
where the Kronecker delta function, ${{\delta _{\omega_{\mathrm{v}}} (\Delta) }}$ with $\Delta  = \omega_{\mathrm{v}}  - {\mathbf{m}} \cdot {\bm \omega} ' + {\mathbf{n}} \cdot {\bm \omega} $, imposes the energy conservation condition. The initial distribution of the phonon modes is denoted by $P_{\rm in}({\bf n})$, which will be taken to be a thermal distribution. Our goal in this work is to explain how a quantum (optical) simulator can be constructed to efficiently sample the FCP at finite temperature. See, for example, Dierksen and Grimme~\cite{Dierksen2005} for the computational difficulties in the evaluation of the FCP. 

\subsection*{Scattershot sampling for thermal states}
We now address the problem of thermal state preparation in Gaussian Boson Sampling, which is relevant in the thermal extension of the sampling problem~\cite{huh2015} for vibronic transitions in molecular spectroscopy. Boson Sampling with thermal input states, as an instance of Gaussian Boson Sampling, has been considered~\cite{rahimi2015}, where it is shown that the distribution can be simulated by a classical computer efficiently. As a result, such a problem belongs to the complexity class $\rm BPP^{NP}$, which is believed to be less complex than the counting problems in the complexity class $\#$P. 

Instead of sampling the thermal distribution, our approach starts with a purification of the mixed initial states, which is a standard method for studying thermo-field dynamics~\cite{Mann1989}, to prepare and identify the thermally excited Fock states given in the Boltzmann distribution. 
Specifically, we extend the idea of Scattershot Boson Sampling~\cite{Lund2014} for the problem of sampling thermalized Bosons. The key idea of Scattershot Boson Sampling is to send half of entangled photons through the optical network, followed by a post-selection for projecting out the single-photon states at the end,
\begin{equation}
P(\mathbf{m},\mathbf{n})=\tr[\hat{O}_{\mathrm{G}}\ket{\mathbf{n}}\bra{\mathbf{n}}\hat{\rho}_{\mathrm{th}}\ket{\mathbf{n}}\bra{\mathbf{n}}\hat{O}_{\mathrm{G}}^{\dagger}\ket{\mathbf{m}}\bra{\mathbf{m}}] \ .
\end{equation}
The main purpose is to overcome the experimental difficulty of preparing single-photon states required in Boson Sampling.



In order to extend the idea of Scattershot Boson Sampling for thermal initial states, we first consider the purification of every thermal state with ancillary modes, i.e.,
$\ket{\mathbf{0}
(\boldsymbol{\beta})}=\sum_{\mathbf{n}=\mathbf{0}}^{\boldsymbol{\infty}}\sqrt{\bra{\mathbf{n}}\hat{\rho}_{\mathrm{th
}}\ket{\mathbf{n}}}\ket{\mathbf{n}}\otimes\ket{\mathbf{n}}_{\mathrm{B}}$,
where the ancillary Hilbert space 'B' has been introduced. Note that the original thermal state $\hat{\rho}_{\mathrm{th}}$ can be obtained after tracing away the ancillary modes, i.e., $\hat{\rho}_{\mathrm{th}}=\tr_{\mathrm{B}}[\ket{\mathbf{0}(\boldsymbol{\beta})}\bra{\mathbf{0}
(\boldsymbol{\beta})}]$, where $\ket{\mathbf{0}(\boldsymbol{\beta})}=\hat{V}
(\boldsymbol{\beta})\ket{\mathbf{0}}\otimes\ket{\mathbf{0}}_{\mathrm{B}}$. The oeprator $\hat{V}(\boldsymbol{\beta})$ is a product 
of two-mode squeezing operators that 
$\hat{V}(\boldsymbol{\beta})=\bigotimes_{k=1}^{M}\exp(\theta_{k}
(\hat{a}_{k}^{\dagger}\hat{b}_{k}^{\dagger}-\hat{a}_{k}\hat{b}_{k})/2)$,
where 
$\tanh(\theta_{k}/2)=\mathrm{e}^{-\beta_{k}\hbar\omega_{k}/2}=\sqrt{\bar{n}_{k}/(\bar{n}_{k}+1)}$,
$\hat{b}_{k}$ and $\hat{b}_{k}^{\dagger}$ are the annihilation and creation operators of the ancillary modes, and $\tr$ and $\tr_{\mathrm{B}}$ trace over the original and ancillary Hilbert spaces, respectively. Here ${{\bar n}_k} = 1/({e^{\beta_k \hbar \omega_k }} - 1)$ is the mean quantum number of the $k$-th mode. Consequently, the sampling problem involving initial sampling of Fock states can be transformed into a problem involving post-selection only, i.e., 
\begin{align}
&P(\mathbf{m},\mathbf{n})=\tr\tr_{\mathrm{B}}\Big[\hat{O}_{\mathrm{G}} \ \hat{V}(\boldsymbol{\beta})\ket{\mathbf{0}}\bra{\mathbf{0}}\otimes\ket{\mathbf{0}}_\mathrm{B}\mathrm{\tensor[_B]{\bra{\mathbf{0}}}{}}\hat{V}(\boldsymbol{\beta})^{\dagger}\hat{O}_{\mathrm{G}}^{\dagger} \ 
 \nonumber\\
&~~~~~~~~~~~~~~~~~~~~~~~~~~~~\ket{\mathbf{m}}\bra{\mathbf{m}}\otimes\ket{\mathbf{n}}_{\mathrm{B}}\mathrm{\tensor[_B]{\bra{\mathbf{n}}}{}} \Big]  .
\label{eq:proboperator}
\end{align}

There are two major differences between Scattershot Boson Sampling~\cite{Lund2014} and our approach. The ancillary modes in Scattershot Boson Sampling are not sent to an optical network and only the measurement results involving single-photon detection are relevant. On the other hand, in our case, 
all the ancillary modes are involved in the optical network in general, and all the measurement outcomes are relevant for the sampling problem. 
Furthermore, the randomized input Fock states are generated with $M$ two-mode squeezed vacuum state that Eq.~\eqref{eq:proboperator} is reduced to the Scattershot Boson Sampling by Lund and coworkers~\cite{Lund2014} when the Gaussian operator includes only the rotation operator, i.e. $\hat{O}_{\mathrm{G}}=\hat{R}_{U}$. In this sense, the Scattershot Boson Sampling is a special instance of Vibronic Boson Sampling.
In the following, we shall show that the mode correlation created by the two-mode squeezing operation ($\hat{V}$) can be eliminated through a Bogoliubov transformation.

\subsection*{Gaussian decorrelation}
To get started, let us define a new operator, 
\begin{equation}
\hat{U}(\boldsymbol{\beta}) \equiv \hat{O}_{\mathrm{G}} \ \hat{V}(\boldsymbol{\beta}) \ ,
\end{equation}
for the Scattershot-fashion probability distribution in Eq.~\ref{eq:proboperator}.
The action of $\hat{U}(\boldsymbol{\beta})$ of the general Gaussian operator is defined for the collective Boson creation operator vector ($\mathbf{c}^{\dagger}=((\mathbf{a}^{\dagger})^{\mathrm{t}},(\mathbf{b}^{\dagger})^{\mathrm{t}})^{\mathrm{t}}$) of the extended space, i.e. 
\begin{equation}
\hat{\mathbf{c}}^{'\dagger}=\hat{U}(\boldsymbol{\beta})^{\dagger}\hat{\mathbf{c}}^{\dagger}\hat{U}(\boldsymbol{\beta})=\mathcal{X}\hat{\mathbf{c}}+\mathcal{Y}\hat{\mathbf{c}}^{\dagger}+\boldsymbol{\gamma} \ ,
\end{equation}
which is obtained by applying the products of two-mode squeezing operators $\hat{V}(\boldsymbol{\beta})$ to 
$\hat{\mathbf{a}}^{'\dagger}$ in Eq.~\ref{eq:genBT} and to $\mathbf{b}^{\dagger}$. 
The resulting parameters are 
\begin{align}
\mathcal{X}=
\begin{pmatrix}
XF & YG\\
G & \mathrm{diag}(\mathbf{0})
\end{pmatrix}, 
\mathcal{Y}=
\begin{pmatrix}
YF & XG\\
\mathrm{diag}(\mathbf{0}) & F
\end{pmatrix}, 
\boldsymbol{\gamma}=
\begin{pmatrix}
\mathbf{z}\\
\mathbf{0}
\end{pmatrix},
\end{align}
and the hyperbolic matrices 
are defined as 
$F=\mathrm{diag}(\sqrt{\bar{n}_{1}+1},\ldots,\sqrt{\bar{n}_{M}+1})$ and  
$G=\mathrm{diag}(\sqrt{\bar{n}_{1}},\ldots,\sqrt{\bar{n}_{M}})$. 

Using the $2M$-dimensional Bogoliubov relation for $\hat{\mathbf{c}}^{'\dagger}$, one can convert 
the Gaussian Boson Sampling with thermal states into the Gaussian Boson Sampling with squeezed coherent ($\boldsymbol{\gamma}\ne\mathbf{0}$)
or vacuum states ($\boldsymbol{\gamma}=\mathbf{0}$) as the input states to the linear photon network (Fig.~\ref{fig:pictorial}).  
We can achieve this goal by means of the SVD of the matrices~\cite{Braunstein2005,Cariolaro2016},  
$\mathcal{X}=\mathcal{C}_{\mathrm{L}}\sinh(\mathcal{S})\mathcal{C}_{\mathrm{R}}^{\mathrm{t}}$ and $\mathcal{Y}=\mathcal{C}_{\mathrm{L}}\cosh(\mathcal{S})\mathcal{C}_{\mathrm{R}}^{\dagger}$ \ ,
where $\mathcal{C}_{\mathrm{L}}$ and $\mathcal{C}_{\mathrm{R}}$ are the unitary matrices, and $\sinh(\mathcal{S})$ and $\cosh(\mathcal{S})$ are diagonal matrices with singular value entries. $\mathcal{S}=\diag(s_{1},\ldots,s_{2M})$ is a diagonal matrix with real values, which correspond to the squeezing parameters.

As a result, the unitary operator in the extended Hilbert space, which is going to be projected on a vacuum state, is decomposed as
\begin{align}
\hat{U}(\boldsymbol{\beta})=\hat{D}_{\boldsymbol{\gamma}^{*}}\hat{R}_{\mathcal{C}_{\mathrm{L}}}\hat{S}_{\mathcal{S}}\hat{R}_{\mathcal{C}_{\mathrm{R}}^{\dagger}}=
\hat{R}_{\mathcal{C}_{\mathrm{L}}}\hat{S}_{\mathcal{S}}\hat{R}_{\mathcal{C}_{\mathrm{R}}^{\dagger}}\hat{D}_{\boldsymbol{\gamma}'} \, ,
\label{eq:decomposition}
\end{align}
The displacement parameter vector after moving the displacement operator from left end to the right end in Eq.~\ref{eq:decomposition}, $\boldsymbol{\gamma}'=\boldsymbol{\gamma}'_{\mathrm{R}}+\mathrm{i}\boldsymbol{\gamma}'_{\mathrm{I}}$, can be calculated by, 
\begin{align}
\begin{pmatrix}
\boldsymbol{\gamma}'_{\mathrm{R}}\\
\boldsymbol{\gamma}'_{\mathrm{I}}
\end{pmatrix}
=
\begin{pmatrix}
\mathcal{X}_{\mathrm{R}}+\mathcal{Y}_{\mathrm{R}} & -\mathcal{X}_{\mathrm{I}}+\mathcal{Y}_{\mathrm{I}}\\
\mathcal{X}_{\mathrm{I}}+\mathcal{Y}_{\mathrm{I}} & \mathcal{X}_{\mathrm{R}}-\mathcal{Y}_{\mathrm{R}}
\end{pmatrix}^{-1}
\begin{pmatrix}
\boldsymbol{\gamma}_{\mathrm{R}}\\
\boldsymbol{\gamma}_{\mathrm{I}}
\end{pmatrix} ,
\label{eq:leqrealimag}
\end{align}
where $\boldsymbol{\gamma}=\boldsymbol{\gamma}_{\mathrm{R}}+\mathrm{i}\boldsymbol{\gamma}_{\mathrm{I}}$, $\mathcal{X}=\mathcal{X}_{\mathrm{R}}+\mathrm{i}\mathcal{X}_{\mathrm{I}}$ and $\mathcal{Y}=\mathcal{Y}_{\mathrm{R}}+\mathrm{i}\mathcal{Y}_{\mathrm{I}}$. 
This linear relation between the displacement parameter vectors for the second equality in Eq.~\ref{eq:decomposition} can be found by applying the two set of sequential operators in Eq.~\ref{eq:decomposition} and comparing the resulting parameter vectors. See Methods for the derivation.  

Finally, the operator (the second equality) in Eq.~\ref{eq:decomposition} can be implemented in quantum optical device via preparing the $2M$-dimensional single-mode squeezed coherent states and passing the squeezed coherent states, 
\begin{equation}
\hat{S}_{\mathcal{S}}\hat{R}_{\mathcal{C}_{\mathrm{R}}^{\dagger}}\hat{D}_{\boldsymbol{\gamma}'}\ket{\mathbf{0}} \equiv \hat{S}_{\mathcal{S}}\vert \boldsymbol{\gamma}''\rangle 
=\bigotimes_{k=1}^{2M}\hat{S}_{s_{k}}\vert \gamma_{k}''\rangle\ ,
\label{eq:twomsinglemode}
\end{equation}
through the linear optical network ($\hat{R}_{\mathcal{C}_{\mathrm{L}}}$). The coherent state parameter vector $\boldsymbol{\gamma}''$ is identified as a rotation of $\boldsymbol{\gamma}'$, i.e. $\boldsymbol{\gamma}''=C_{\mathrm{R}}^{\mathrm{t}}\boldsymbol{\gamma}'$.  
The correlated squeezed thermal state ($\hat{S}\hat{R}\rho_{\mathrm{th}}\hat{R}^{\dagger}\hat{S}^{\dagger}$) is replaced with the $2M$-dimensional single-mode squeezed coherent states. 
In Fig.~\ref{fig:pictorial}, the quantum optical unraveling of the thermal state is depicted. This can be applied to any thermal state involved problem, e.g. thermal state Boson Sampling and molecular vibronic spectroscopy at finite temperature~\cite{berger:1998,Huh2011a}. The connection to the molecular problem is given in Methods.   
\subsection*{Molecular vibronic spectroscopy at finite temperature}
Here, we write the molecular vibronic Franck-Condon Profile (FCP) at finite temperature $T$ ($\beta=\beta_{1}=\cdots=\beta_{M}=1/k_{\mathbf{B}}T$) in the extended Hilbert space as follows 
\begin{align}
&\mathrm{FCP}(\omega_{\mathrm{v}})= \label{eq:FCP} \sum_{\mathbf{m},\mathbf{n}}^{\boldsymbol{\infty}}P_{\mathbf{m}\mathbf{n}}(\beta)\delta(\omega_{\mathrm{v}}-(\mathbf{m}\cdot\boldsymbol{\omega}'-\mathbf{n}\cdot\boldsymbol{\omega})) \, ,\\
&P_{\mathbf{m}\mathbf{n}}(\beta)=\mathrm{tr}\mathrm{tr}_{\mathrm{B}}\Big[\hat{U}_{\mathrm{Dok}}\ket{\mathbf{0}(\beta)}\bra{\mathbf{0}(\beta)}\hat{U}_{\mathrm{Dok}}^{\dagger} \nonumber\\
&~~~~~~~~~~~~~~~~~~~~~~~\ket{\mathbf{m}}\bra{\mathbf{m}}\otimes\ket{\mathbf{n}}_{\mathrm{B}}\mathrm{\tensor[_B]{\bra{\mathbf{n}}}{}} \Big] .
\end{align}
Unlike the zero temperature case ($\omega_{\mathrm{v}}=\mathbf{m}\cdot\boldsymbol{\omega}'$)~\cite{huh2015}, the transition frequency at finite temperature has the negative contribution from the initial quantum oscillators that $\omega_{\mathrm{v}}=\mathbf{m}\cdot\boldsymbol{\omega}'-\mathbf{n}\cdot\boldsymbol{\omega}$.  

The Doktorov operator $\hat{U}_{\mathrm{Dok}}(=\hat{U}_{\mathrm{D}})$~\cite{doktorov:1977,huh2015} for the molecular scattering ($\hat{O}$) is defined as 
\begin{align}
\hat{U}_{\mathrm{Dok}}&=\hat{D}_{\boldsymbol{\delta}/\sqrt{2}}
\hat{S}_{\ln\Omega'}\hat{R}_{U}\hat{S}_{\ln\Omega}^{\dagger} \, ,
\label{eq:Doktorov}
\end{align}
where this applies to the optical mode $M$. The multidimensional Bogoliubov transformation of $\mathbf{c}^{\dagger}$ resulting by $\hat{U}=\hat{U}_{\mathrm{Dok}}\hat{V}(\beta)$ is defined with the following Bogoliubov matrices and displacement vector,  

\begin{align}
&\mathcal{X}=
\begin{pmatrix}
\frac{1}{2}\left(J-(J^{\mathrm{t}})^{-1}\right)F && \frac{1}{2}\left(J+(J^{\mathrm{t}})^{-1}\right)G\\
G && 0
\end{pmatrix},
\\
&\mathcal{Y}=
\begin{pmatrix}
\frac{1}{2}\left(J+(J^{\mathrm{t}})^{-1}\right)F && \frac{1}{2}\left(J-(J^{\mathrm{t}})^{-1}\right)G\\
0 && F
\end{pmatrix},
\\
&\boldsymbol{\gamma}=
\begin{pmatrix}
\frac{1}{\sqrt{2}}\boldsymbol{\delta}\\
\mathbf{0}
\end{pmatrix},
    \label{eq:duschinskya}
\end{align}
where $\boldsymbol{\delta}$ is a molecular displacement vector, and $\mathbf{J}$ is defined as follows
\begin{align}
&J=\Omega'U\Omega^{-1}, \nonumber \\
&\Omega'=\mathrm{diag}(\sqrt{\omega_{1}'},\ldots,\sqrt{\omega_{N}'}), \quad 
\Omega=\mathrm{diag}(\sqrt{\omega_{1}},\ldots,\sqrt{\omega_{N}}) \, ,
\label{eq:parameters}
\end{align}
with the Duschinsky unitary rotation matrix $U$~\cite{jankowiak:2007,Huh2011a}. 
The FCP at finite temperature can be implemented with linear optical network as depicted in Fig. 1c with the 
sing-mode squeezed coherent (vacuum) states in Eq.~\ref{eq:twomsinglemode}. 

\subsubsection*{Numerical Example}

\begin{figure}[htb]
\begin{center}
\includegraphics[width=0.5\columnwidth]{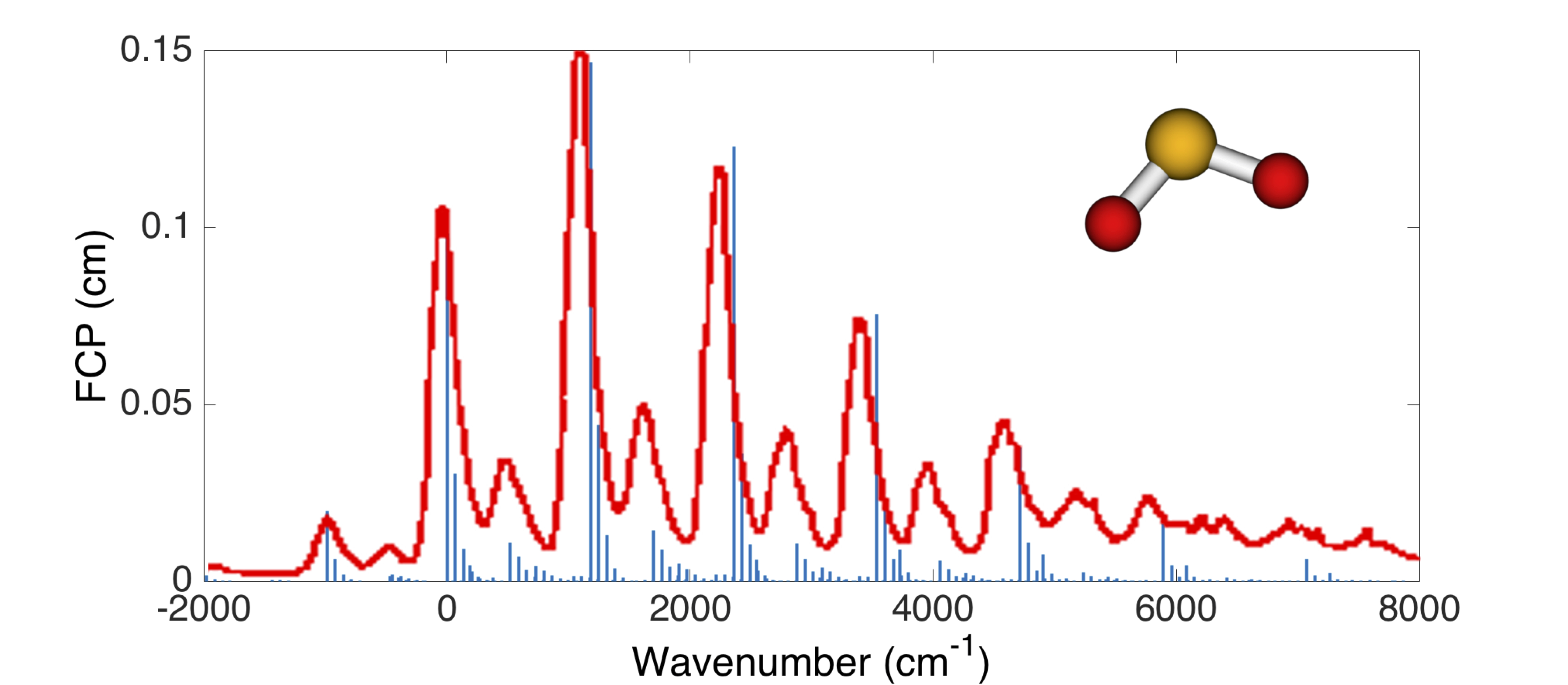}
\caption{Franck-Condon profile of the photoelectron spectroscopy of sulfur dioxide anion (SO$_{2}^{-}$ $\rightarrow$ SO$_{2}$)~\cite{Lee2009} at finite temperature (650 K). The sticks are calculated with classical computer and the red curves are taken from the molecular spectroscopic experiment~\cite{nimlos1986}. 
}\label{fig:so2anspec}
\end{center}
\end{figure}

\begin{figure}[htb]
\begin{center}
\includegraphics[width=0.5\columnwidth]{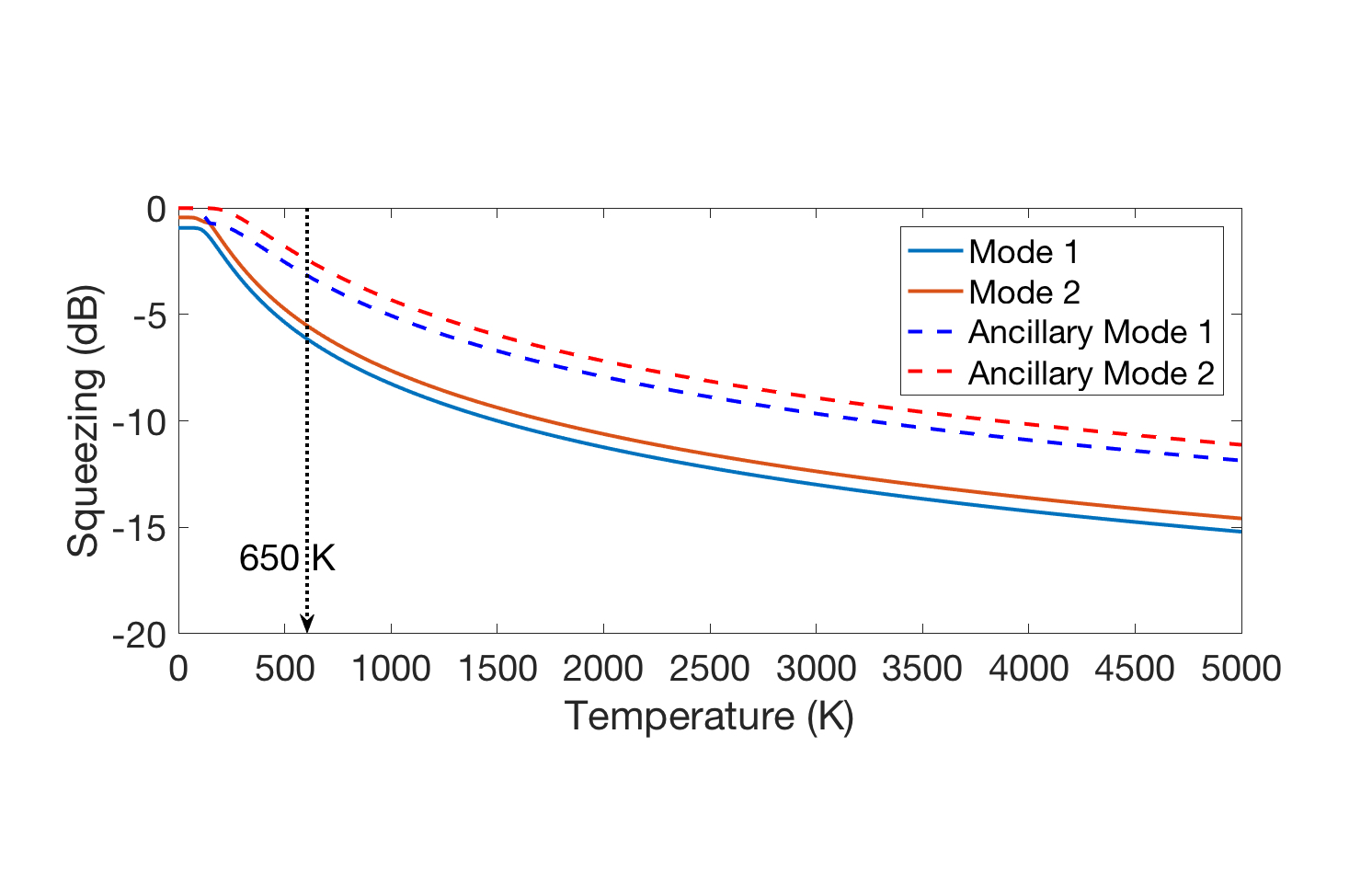}
\caption{Squeezing parameters of the photoelectron spectroscopy of sulfur dioxide anion (SO$_{2}^{-}$ $\rightarrow$ SO$_{2}$)~\cite{Lee2009} at different temperatures. The squeezing parameters in dB is obtained as $10\log_{10}(\mathrm{e}^{-2s_{k}})$, $s_k$ is a squeezing parameter and one of the diagonal matrix elements of $\mathcal{S}$.  
}\label{fig:so2ansqueezing}
\end{center}
\end{figure}

We present in Fig.~\ref{fig:so2anspec}, the photoelectron spectrum of sulfur dioxide anion (SO$_{2}^{-}$ $\rightarrow$ SO$_{2}$)~\cite{Lee2009} at finite temperature (650 K)~\cite{nimlos1986,Lee2009} as an example for the optical setup in Fig.~\ref{fig:pictorial}.   
The FCP at finite temperature in Eq.~\ref{eq:FCP} is computed with a classical computer for this two-dimensional example and presented as sticks in the figure. The corresponding molecular spectroscopic curve from Ref.~\cite{nimlos1986} is overlaid in red.    
The molecular specific parameters are given in Methods. Unlike the FCP at zero temperature, the FCP at finite temperature has peaks in the negative frequency domain due to the thermal excitation of the molecule. 
By using Eq.~\ref{eq:decomposition}, the quantum optical apparatus can be constructed 
for the quantum simulation as depicted in Fig.~\ref{fig:pictorial}. In Fig.~\ref{fig:so2ansqueezing}, the squeezing parameters in dB of the optical modes including the ancillary modes at varying temperature are shown as solid and dashed lines, respectively. The magnitudes of the squeezing parameters start from below 1 dB at 0 K, and then they increase as temperature increases because the two-mode squeezing parameter $\theta_{k}$ increases. 
The squeezing parameters at 650 K corresponding to the spectrum in Fig.~\ref{fig:so2anspec} are below 6.5 dB as indicated in Fig.~\ref{fig:so2ansqueezing}.

\begin{figure}[htb]
\begin{center}
\includegraphics[width=0.45 \textwidth]{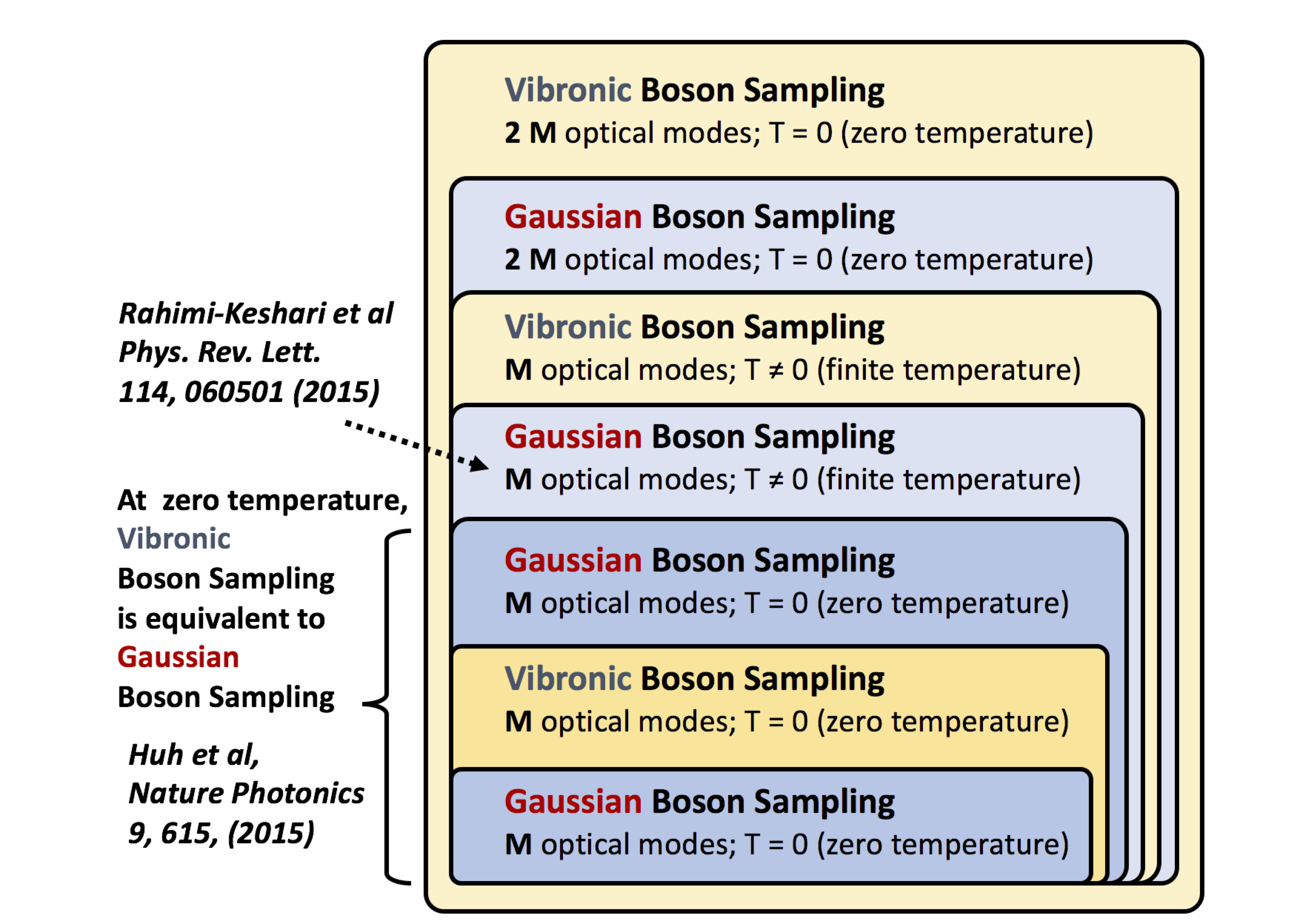}

\caption{Hierarchy of Gaussian Boson Sampling. The Hierarchy of Gaussian Boson Samplings are structured according to the temperature and the size of the problem. For the same temperature and same number of modes, Gaussian Boson Sampling~\cite{rahimi2015} is always a special case  of Vibronic Boson Sampling, except at zero temperature, the two classes are the same as proved in Ref.~\cite{huh2015}. In this work, we prove that Vibronic Boson Sampling at finite temperatures can be transformed into Gaussian Boson Sampling by doubling the optical modes.
}\label{fig:bshier}
\end{center}
\end{figure} 

\section*{Discussion}
Our discussion can now be summarized by the following hierarchical structure depicted in Fig.~\ref{fig:bshier} for the relations between the various versions of Gaussian Boson Samplings. 
First, the $M$-dimensional Gaussian Boson Sampling with independent Boson modes at zero temperature ($M$-GBS(T=0)) is a special case of the $M$-dimensional Vibronic Boson Sampling at zero temperature ($M$-VBS(T=0)), which implies $M$-GBS(T=0)$\subseteq$$M$-VBS(T=0). On the other hand, $M$-VBS(T=0) is always reduced to $M$-GBS(T=0) because $\hat{R}\ket{\mathbf{0}}\bra{\mathbf{0}}\hat{R}^{\dagger}=\ket{\mathbf{0}}\bra{\mathbf{0}}$, i.e. $M$-VBS(T=0)$\subseteq$$M$-GBS(T=0). Thus, the Vibonic Boson Sampling introduced by Huh et al.~\cite{huh2015} and the Gaussian Boson Sampling by Rahimi-Keshari et al.~\cite{rahimi2015} are equivalent at zero temperature. Therefore, $M$-GBS(T=0)$\equiv$$M$-VBS(T=0)$\subseteq$$M$-GBS(T$\neq$0). When the mean quantum numbers of all modes ($\bar{n}_{k}$) are equal or the identity rotation ($\hat{R}_{I}$) is introduced the thermal state is invariant to the rotation leaving no correlation between the modes, i.e. $\hat{R}\hat{\rho}_{\mathrm{th}}\hat{R}^{\dagger}=\hat{\rho}_{\mathrm{th}}$. The Vibronic Boson Sampling with correlated Bosons at finite temperature includes the Gaussian Boson Sampling with independent Bosons at zero Kelvin that it means $M$-GBS(T$\neq$0)$\subseteq$$M$-VBS(T$\neq$0).  
$M$-VBS(T$\neq$0) can always to be transformed to the Gaussian Boson Sampling at zero temperature with the uncorrelated Boson modes ($2M$-GBS(T=0)) via the extended Hilbert space approach in Eq.~\ref{eq:proboperator}. 
Moreover, the extended Hilbert space approach can transform the correlated thermal state into 
the single-mode squeezed states as in Eq.~\ref{eq:twomsinglemode}. 
This means that $M$-VBS(T$\neq$0)$\subseteq$$2M$-GBS(T=0). With the same argument above for the $M$-dimensional case, $2M$-GBS(T=0)$\subseteq$$2M$-VBS(T=0). (See also Methods for the relation in the phase space representation of the Gaussian state.) 
Scattershot Boson Sampling, with post-selection, is equivalent to the original Boson Sampling and the Scattershot Boson Sampling is a special instance of the Vibronic Boson Sampling, therefore, 
"original Boson Sampling $\subseteq$ Scattershot Boson Sampling $\subseteq$ Vibronic Boson Sampling" in the extended Hilbert space. 

In closing, we studied the problem of generalizing Gaussian Boson Sampling with initial correlation of input Bosons. This problem is relevant to the molecular spectroscopy problem at finite temperature; and we show this with a specific example of the photoelectron process of SO$_{2}^{-}$ at 650 K. We employed a multidimensional Bogoliubov transformation together with an extended Hilbert space to de-correlate the Gaussian input state. Furthermore, we present a hierarchy for clarifying the relationships between various types of Gaussian Boson Samplings.  Finally, our results imply an explicit scattershot approach for quantum-optical realization of sampling thermal Bosons without the need of an explicit preparation procedure for the Boltzmann distribution. 

\section*{Methods}

\subsection*{Quantum optical operators}
A shorthand notation for the Boson operator vector $\hat{\mathbf{x}}$ has been used, \emph{i.e.}
$\hat{A} \ \hat{\mathbf{x}} \ \hat{B} \equiv (\hat{A}\hat{x}_{1}\hat{B},\ldots,\hat{A}\hat{x}_{M}\hat{B})^{\mathrm{t}}$ in the paper.
The displacement, squeezing and rotation operators are defined as follow, respectively, 
\begin{equation}
\hat{D}_{\boldsymbol{\alpha}}=\exp(\boldsymbol{\alpha}^{\mathrm{t}} \hat{\mathbf{a}}^{\dagger}-\boldsymbol{\alpha}^{\dagger}\hat{\mathbf{a}}) \ ,
\end{equation}
\begin{equation}
\hat{S}_{\Sigma}=\exp(((\hat{\mathbf{a}}^{\dagger})^{\mathrm{t}}\Sigma\hat{\mathbf{a}}^{\dagger}-\hat{\mathbf{a}}^{\mathrm{t}}\Sigma\hat{\mathbf{a}})/2) \ ,
\end{equation}
and 
\begin{equation}
\hat{R}_{U}=\exp((\hat{\mathbf{a}}^{\dagger})^{\mathrm{t}}\ln U^{*}\hat{\mathbf{a}}) \ .
\end{equation}
Here $\boldsymbol{\alpha}$ is a $M$-dimensional coherent state phase vector, $\Sigma$ is a diagonal matrix with real diagonal entries of squeezing parameters, and $U$ is a $M\times M$ unitary matrix. 
The action of quantum optical operators to the Boson creation operator $\mathbf{\hat{a}}^{\dagger}$ are defined as, see also Ref.~\cite{Ma1990}, 
\begin{align}
&\hat{D}_{\boldsymbol{\alpha}}^{\dagger}\mathbf{\hat{a}}^{\dagger}\hat{D}_{\boldsymbol{\alpha}}=\mathbf{\hat{a}}^{\dagger}+\boldsymbol{\alpha}^{*}, 
\label{eq:ql1}
\\
&\hat{S}_{\Sigma}^{\dagger}\mathbf{\hat{a}}^{\dagger}\hat{S}_{\Sigma}=
\sinh(\Sigma)\mathbf{\hat{a}}
+\cosh(\Sigma)\mathbf{\hat{a}}^{\dagger}, 
\label{eq:ql2}
\\
&\hat{R}_{U}^{\dagger}\hat{\mathbf{a}}^{\dagger}
                                \hat{R}_{U}
                            = U\hat{\mathbf{a}}^{\dagger} \,  .
\label{eq:ql3}
\end{align}

The action of the squeezing operator $\hat{V}(\boldsymbol{\beta})=\bigotimes_{k=1}^{M}\exp(\theta_{k}
(\hat{a}_{k}^{\dagger}\hat{b}_{k}^{\dagger}-\hat{a}_{k}\hat{b}_{k})/2)$ is given by,
\begin{align}
  \hat V{(\bm{\beta} )^\dag }{{\hat a}_k}\hat V(\bm{\beta} ) & = \cosh ({\theta _k}/2) \ {{\hat a}_k} + \sinh ({\theta _k}/2) \ \hat b_k^\dag  \hfill \ , \\
  \hat V{(\bm{\beta} )^\dag }{{\hat b}_k}\hat V(\bm{\beta} ) & = \sinh (\theta_k/2) \ \hat a_k^\dag  + \cosh ({\theta _k}/2) \ {{\hat b}_k} \hfill \ .
\end{align}

\subsection*{Derivation of Eq.~\ref{eq:leqrealimag}} 
We rewrite $\hat{U}$ as   $\hat{U}=\hat{D}_{\boldsymbol{\gamma}^{*}}\hat{U}_{0}=\hat{U}_{0}\hat{D}_{\boldsymbol{\gamma}'}$, where  $\hat{U}_{0}=\hat{R}_{\mathcal{C}_{\mathrm{L}}}\hat{S}_{\mathcal{S}}\hat{R}_{\mathcal{C}_{\mathrm{R}}^{\dagger}}$. 
The action of $\hat{U}_{0}$ to $\mathbf{\hat{c}}^{\dagger}$ can be found easily with the identities in Eqs.~\ref{eq:ql1},~\ref{eq:ql2} and~\ref{eq:ql3}, as follows
\begin{align}
\hat{U}_{0}^{\dagger}\mathbf{\hat{c}}^{\dagger}\hat{U}_{0}=\mathcal{X}\hat{\mathbf{c}}+\mathcal{Y}\hat{\mathbf{c}}^{\dagger}, 
\end{align}
where 
$\mathcal{X}=\mathcal{C}_{\mathrm{L}}\sinh(\mathcal{S})\mathcal{C}_{\mathrm{R}}^{\mathrm{t}}, \quad {\rm and}$
$\mathcal{Y}=\mathcal{C}_{\mathrm{L}}\cosh(\mathcal{S})\mathcal{C}_{\mathrm{R}}^{\dagger}$~\cite{Braunstein2005,Cariolaro2016}. 
Now we work out $\hat{U}^{\dagger}\mathbf{\hat{c}}^{\dagger}\hat{U}$ in the two different ways, i.e. for $\hat{U}=\hat{D}_{\boldsymbol{\gamma}^{*}}\hat{U}_{0}$ and $\hat{U}=\hat{U}_{0}\hat{D}_{\boldsymbol{\gamma}'}$ . 
The resulting linear transforms are as follow, 
\begin{align}
\hat{U}_{0}^{\dagger}\hat{D}_{\boldsymbol{\gamma}^{*}}^{\dagger}\mathbf{\hat{c}}^{\dagger}\hat{D}_{\boldsymbol{\gamma}^{*}}\hat{U}_{0}&=\hat{U}_{0}^{\dagger}(\mathbf{\hat{c}}^{\dagger}+\boldsymbol{\gamma})\hat{U}_{0} \nonumber \\
&=\mathcal{X}\hat{\mathbf{c}}+\mathcal{Y}\hat{\mathbf{c}}^{\dagger}+\boldsymbol{\gamma} ,
\label{eq:leftop}
\end{align}

\begin{align}
\hat{D}_{\boldsymbol{\gamma}'}^{\dagger}\hat{U}_{0}^{\dagger}\mathbf{\hat{c}}^{\dagger}\hat{U}_{0}\hat{D}_{\boldsymbol{\gamma}'}&= \hat{D}_{\boldsymbol{\gamma}'}^{\dagger}(\mathcal{X}\hat{\mathbf{c}}+\mathcal{Y}\hat{\mathbf{c}}^{\dagger}) \hat{D}_{\boldsymbol{\gamma}'}\nonumber \\
&=\mathcal{X}\hat{\mathbf{c}}+\mathcal{Y}\hat{\mathbf{c}}^{\dagger}+\mathcal{X}\boldsymbol{\gamma}'+\mathcal{Y}\boldsymbol{\gamma}^{'*} .
\label{eq:rightop}
\end{align}
By comparing Eqs.~\ref{eq:leftop} and \ref{eq:rightop}, we have $\boldsymbol{\gamma}=\mathcal{X}\boldsymbol{\gamma}'+\mathcal{Y}\boldsymbol{\gamma}^{'*}$. After separating the real and imaginary parts of this linear equation we can derive Eq.~\ref{eq:leqrealimag}. 

\subsection*{Molecular parameters}
Molecular parameters for SO$_{2}$ molecule are taken from Ref.~\cite{Lee2009}. 
\begin{align}
&\boldsymbol{\omega}=(989.5, 451.4)^{\mathrm{t}}, 
\boldsymbol{\omega}=(1178.4, 518.9)^{\mathrm{t}}\\
&U=\begin{pmatrix}
0.9979 &   0.0646\\
   -0.0646 &   0.9979
\end{pmatrix},
\boldsymbol{\delta}=(-1.8830, 0.4551)^{\mathrm{t}} .
\end{align}

The parameters at 650 K for the quantum simulation corresponding to Fig.~\ref{fig:so2anspec}. 

\begin{align}
&\mathcal{C}_{\mathrm{L}}=
\begin{pmatrix}
    0.0963 &   0.0114 &   0.7505 &  -0.6537\\
    0.7297 &   0.6789 &  -0.0738 &   0.0346\\
    0.0169 &   0.0147 &   0.6553 &   0.7550\\
    0.6767 &  -0.7340 &  -0.0435 &   0.0369
\end{pmatrix},\\
&\mathcal{S}=
\mathrm{diag}(0.7419, 0.6701, 0.3932, 0.3080),\\
&\mathcal{C}_{\mathrm{R}}=
\begin{pmatrix}
    0.0386 &   0.0299 &   0.7448 &   0.6656\\
    0.7197 &  -0.6937 &  -0.0230 &   0.0151\\
    0.0205 &  -0.0171 &   0.6659 &  -0.7456\\
    0.6929 &   0.7194 &  -0.0373 &  -0.0308\\
\end{pmatrix}.
\end{align}

\subsection*{Gaussian state in phase space}
Here we present a multivariate Gaussian state~\cite{Weedbrook2012,Adesso2014} in phase space regarding to Husimi function~\cite{rahimi2015} in connection to the multidimensional Bogoliubov transformation of the thermal state, which is used in this paper. 
The Husimi $\mathcal{Q}$ function of the most general Gaussian state ($\hat{\rho}_{\mathrm{G}}$) is given as 
\begin{align}
\mathcal{Q}(\boldsymbol{\alpha})&=\frac{1}{\pi^{M}}\langle\boldsymbol{\alpha}\vert
\hat{\rho}_{\mathrm{in}}\vert\boldsymbol{\alpha}\rangle,\\
&=\frac{\exp[-\frac{1}{2}(\overrightarrow{\boldsymbol{\alpha}}-\overrightarrow{\mathbf{z}})^{\mathrm{t}}
L^{*}(V+I)^{-1}L^{\dagger}(\overrightarrow{\boldsymbol{\alpha}}-\overrightarrow{\mathbf{z}})]}{(2\pi)^{M}\sqrt{\det(V+I)}},
\label{eq:husimiq}
\end{align}
where $\overrightarrow{\boldsymbol{\alpha}}=(\boldsymbol{\alpha}^{\mathrm{t}},\boldsymbol{\alpha}^{\dagger})^{\mathrm{t}}$ and $\overrightarrow{\mathbf{z}}=(\mathbf{z}^{\mathrm{t}},\mathbf{z}^{\dagger})^{\mathrm{t}}$. $V$ is a covariance matrix in position-momentum basis~\cite{Weedbrook2012} and $L$ is a basis conversion matrix from the coherent phase basis to the position-momentum basis~\cite{Adesso2014}.   
\begin{align}
L^{\dagger}=
\begin{pmatrix}
I & I\\
-\mathrm{i}I & \mathrm{i}I
\end{pmatrix} ,
\label{eq:conversionmatrix}
\end{align}
such that $(\hat{\mathbf{q}},\hat{\mathbf{p}})^{\mathrm{t}}=L^{\dagger}(\hat{\mathbf{a}},\hat{\mathbf{a}}^{\dagger})^{\mathrm{t}}$. $(\hat{\mathbf{q}},\hat{\mathbf{p}})^{\mathrm{t}}$ is a $M$-dimensional momentum and a $M$-dimensional position operator vectors, respectively. 
We note here Ref.~\cite{Weedbrook2012} and Ref.~\cite{Adesso2014} use different scaling convention in relating the Boson creation and annihilation operators to the position and momentum operators. According to Eq.~\ref{eq:conversionmatrix},  we use the convention of $\hat{q}_{k}=\hat{a}_{k}+\hat{a}^{\dagger}_{k}$ and $\hat{p}_{k}=-\mathrm{i}(\hat{a}_{k}-\hat{a}^{\dagger}_{k})$ as used in Ref.~\cite{Weedbrook2012} to derive Eq.~\ref{eq:husimiq} being consistent with the expression in Ref.~\cite{rahimi2015}. 
The covariance matrix is obtained as~\cite{Adesso2014}, 
\begin{align}
&V=L^{\dagger}W\Xi W^{\mathrm{t}}L^{*},\\
&W=
\begin{pmatrix}
Y^{*}& X^{*}\\
X & Y
\end{pmatrix}, 
\Xi=
\begin{pmatrix}
\mathrm{diag}(\mathbf{0})& \mathrm{diag}(\boldsymbol{\nu})\\
\mathrm{diag}(\boldsymbol{\nu}) & \mathrm{diag}(\mathbf{0})
\end{pmatrix}, 
\end{align}
where $\boldsymbol{\nu}=(2\bar{n}_{1}+1,\ldots,2\bar{n}_{M}+1)$. $\bar{n}_{k}$ is the mean photon number of $k$-th mode that $\bar{n}_{k}=[\exp(\beta_{k}\hbar\omega_{k})-1]^{-1}$.

The probability $P(\mathbf{m})$ is given in an integral form with the Husimi $\mathcal{Q}$ function, accordingly, 
\begin{align}
P(\mathbf{m})=\pi^{-M}\int\mathrm{d}^{2}\boldsymbol{\alpha} \mathcal{Q}(\boldsymbol{\alpha})\mathcal{P}_{\mathbf{m}}(\boldsymbol{\alpha}), 
\end{align}
where $\mathcal{P}_{\mathbf{m}}(\boldsymbol{\alpha})$ is the Glauber-Sudarshan $\mathcal{P}$ function of the number state $\vert\mathbf{m}\rangle\langle\mathbf{m}\vert$~\cite{glauber:1963,sudarshan1963}.

As a special case, when $\mathbf{z}=\mathbf{0}$ and $U_{\mathrm{R}}=I$, the generalized Gaussian state $\hat{\rho}_{\mathrm{G}}$ is reduced to the Gaussian state considered in Ref.~\cite{rahimi2015}. Such that, the corresponding Husimi $\mathcal{Q}$ function is obtained from Eq.~\ref{eq:husimiq}, i.e.
 \begin{align}
 &\mathcal{Q}(\boldsymbol{\alpha})=\frac{\prod_{k=1}^{M}\sqrt{\mu_{k}^{2}-4\lambda_{k}^{2}}}{\pi^{M}}\\
 &\exp\left[\overrightarrow{\boldsymbol{\alpha}}^{\mathrm{t}}
 \begin{pmatrix}
 U_{\mathrm{L}}\diag(\boldsymbol{\lambda})U_{\mathrm{L}}^{\mathrm{t}} &  
 -U_{\mathrm{L}}\diag(\boldsymbol{\mu})U_{\mathrm{L}}^{\dagger}/2\\
  -U_{\mathrm{L}}^{*}\diag(\boldsymbol{\mu})U_{\mathrm{L}}^{\mathrm{t}}/2 & U_{\mathrm{L}}^{*}\diag(\boldsymbol{\lambda})U_{\mathrm{L}}^{\dagger}
 \end{pmatrix}
\overrightarrow{\boldsymbol{\alpha}}
 \right] ,
 \label{eq:husimireduced}
 \end{align}
 where $\lambda_{k}=\frac{1}{2}((V_{x_{k}}+1)^{-1}-(V_{p_{k}}+1)^{-1})$ and $\mu_{k}=(V_{x_{k}}+1)^{-1}+(V_{p_{k}}+1)^{-1}$, and $V_{x_{k}}=(2\bar{n}_{k}+1)\exp(2\sigma_{k})$ and $V_{p_{k}}=(2\bar{n}_{k}+1)\exp(-2\sigma_{k})$.

The Gaussian state used by Lund et al.~\cite{Lund2014} is a special case of Eq. 9 that $\boldsymbol{\gamma}=\mathbf{0}$, $\mathcal{S}=\diag(s_{1},\ldots,s_{2M})$, $\mathcal{C}_{\mathrm{L}}=\diag(U,I)$ and $\mathcal{C}_{\mathrm{R}}^{\dagger}=\diag(I,I)$. Again, this is a special instance of Eq.~\ref{eq:husimireduced} at zero temperature ($\boldsymbol{\mu}=(1,1,\ldots)^{\mathrm{t}}$).


\begin{thebibliography}{10}
\expandafter\ifx\csname url\endcsname\relax
  \def\url#1{\texttt{#1}}\fi
\expandafter\ifx\csname urlprefix\endcsname\relax\def\urlprefix{URL }\fi
\expandafter\ifx\csname doiprefix\endcsname\relax\def\doiprefix{DOI }\fi
\providecommand{\bibinfo}[2]{#2}
\providecommand{\eprint}[2][]{\url{#2}}

\bibitem{Aaronson2011}
\bibinfo{author}{Aaronson, S.} \& \bibinfo{author}{Arkhipov, A.}
\newblock \bibinfo{journal}{\bibinfo{title}{{The computational complexity of
  linear optics}}}.
\newblock {\emph{\JournalTitle{Proceedings of the 43rd annual ACM symposium on
  Theory of computing - STOC '11}}} \bibinfo{pages}{333}
  (\bibinfo{year}{2011}).

\bibitem{Spring2013}
\bibinfo{author}{Spring, J.~B.}, \bibinfo{author}{Metcalf, B.~J.},
  \bibinfo{author}{Humphreys, P.~C.}, \bibinfo{author}{Kolthammer, W.~S.},
  \bibinfo{author}{Jin, X.-M.}, \bibinfo{author}{Barbieri, M.},
  \bibinfo{author}{Datta, A.}, \bibinfo{author}{Thomas-Peter, N.},
  \bibinfo{author}{Langford, N.~K.}, \bibinfo{author}{Kundys, D.},
  \bibinfo{author}{Gates, J.~C.}, \bibinfo{author}{Smith, B.~J.},
  \bibinfo{author}{Smith, P. G.~R.} \& \bibinfo{author}{Walmsley, I.~A.}
\newblock \bibinfo{journal}{\bibinfo{title}{{Boson sampling on a photonic
  chip}}}.
\newblock {\emph{\JournalTitle{Science}}} \textbf{\bibinfo{volume}{339}},
  \bibinfo{pages}{798} (\bibinfo{year}{2013}).

\bibitem{Broome2013}
\bibinfo{author}{Broome, M.~A.}, \bibinfo{author}{Fedrizzi, A.},
  \bibinfo{author}{Rahimi-Keshari, S.}, \bibinfo{author}{Dove, J.},
  \bibinfo{author}{Aaronson, S.}, \bibinfo{author}{Ralph, T.~C.} \&
  \bibinfo{author}{White, A.~G.}
\newblock \bibinfo{journal}{\bibinfo{title}{{Photonic boson sampling in a
  tunable circuit}}}.
\newblock {\emph{\JournalTitle{Science}}} \textbf{\bibinfo{volume}{339}},
  \bibinfo{pages}{794} (\bibinfo{year}{2013}).

\bibitem{Crespi2013}
\bibinfo{author}{Crespi, A.}, \bibinfo{author}{Osellame, R.},
  \bibinfo{author}{Ramponi, R.}, \bibinfo{author}{Brod, D.~J.},
  \bibinfo{author}{Galvao, E.~F.}, \bibinfo{author}{Spagnolo, N.},
  \bibinfo{author}{Vitelli, C.}, \bibinfo{author}{Maiorino, E.},
  \bibinfo{author}{Mataloni, P.} \& \bibinfo{author}{Sciarrino, F.}
\newblock \bibinfo{journal}{\bibinfo{title}{{Integrated multimode
  interferometers with arbitrary designs for photonic boson sampling}}}.
\newblock {\emph{\JournalTitle{Nature Photon.}}} \textbf{\bibinfo{volume}{7}},
  \bibinfo{pages}{545--549} (\bibinfo{year}{2013}).

\bibitem{Tillmann2013}
\bibinfo{author}{Tillmann, M.}, \bibinfo{author}{Dakic, B.},
  \bibinfo{author}{Heilmann, R.}, \bibinfo{author}{Nolte, S.},
  \bibinfo{author}{Szameit, A.} \& \bibinfo{author}{Walther, P.}
\newblock \bibinfo{journal}{\bibinfo{title}{{Experimental boson sampling}}}.
\newblock {\emph{\JournalTitle{Nature Photon.}}} \textbf{\bibinfo{volume}{7}},
  \bibinfo{pages}{540} (\bibinfo{year}{2013}).

\bibitem{Lund2014}
\bibinfo{author}{Lund, A.~P.}, \bibinfo{author}{Laing, A.},
  \bibinfo{author}{Rahimi-Keshari, S.}, \bibinfo{author}{Rudolph, T.},
  \bibinfo{author}{O’Brien, J.~L.} \& \bibinfo{author}{Ralph, T.~C.}
\newblock \bibinfo{journal}{\bibinfo{title}{{Boson Sampling from a Gaussian
  state}}}.
\newblock {\emph{\JournalTitle{Phys. Rev. Lett.}}}
  \textbf{\bibinfo{volume}{113}}, \bibinfo{pages}{100502}
  (\bibinfo{year}{2014}).

\bibitem{Bentivegna2015}
\bibinfo{author}{Bentivegna, M.}, \bibinfo{author}{Spagnolo, N.},
  \bibinfo{author}{Vitelli, C.}, \bibinfo{author}{Flamini, F.},
  \bibinfo{author}{Viggianiello, N.}, \bibinfo{author}{Latmiral, L.},
  \bibinfo{author}{Mataloni, P.}, \bibinfo{author}{Brod, D.~J.},
  \bibinfo{author}{Galvao, E.~F.}, \bibinfo{author}{Crespi, A.},
  \bibinfo{author}{Ramponi, R.}, \bibinfo{author}{Osellame, R.} \&
  \bibinfo{author}{Sciarrino, F.}
\newblock \bibinfo{journal}{\bibinfo{title}{{Experimental scattershot boson
  sampling}}}.
\newblock {\emph{\JournalTitle{Sci. Adv.}}} \textbf{\bibinfo{volume}{1}},
  \bibinfo{pages}{e1400255} (\bibinfo{year}{2015}).

\bibitem{Olson2015}
\bibinfo{author}{Olson, J.~P.}, \bibinfo{author}{Seshadreesan, K.~P.},
  \bibinfo{author}{Motes, K.~R.}, \bibinfo{author}{Rohde, P.~P.} \&
  \bibinfo{author}{Dowling, J.~P.}
\newblock \bibinfo{journal}{\bibinfo{title}{{Sampling arbitrary photon-added or
  photon-subtracted squeezed states is in the same complexity class as boson
  sampling}}}.
\newblock {\emph{\JournalTitle{Phys. Rev. A}}} \textbf{\bibinfo{volume}{91}},
  \bibinfo{pages}{022317} (\bibinfo{year}{2015}).

\bibitem{Seshadreesan2015}
\bibinfo{author}{Seshadreesan, K.~P.}, \bibinfo{author}{Olson, J.~P.},
  \bibinfo{author}{Motes, K.~R.}, \bibinfo{author}{Rohde, P.~P.} \&
  \bibinfo{author}{Dowling, J.~P.}
\newblock \bibinfo{journal}{\bibinfo{title}{{Boson sampling with displaced
  single-photon Fock states versus single-photon-added coherent states: The
  quantum-classical divide and computational-complexity transitions in linear
  optics}}}.
\newblock {\emph{\JournalTitle{Phys. Rev. A}}} \textbf{\bibinfo{volume}{91}},
  \bibinfo{pages}{022334} (\bibinfo{year}{2015}).

\bibitem{Rohde2015}
\bibinfo{author}{Rohde, P.~P.}, \bibinfo{author}{Motes, K.~R.},
  \bibinfo{author}{Knott, P.~A.}, \bibinfo{author}{Fitzsimons, J.},
  \bibinfo{author}{Munro, W.~J.} \& \bibinfo{author}{Dowling, J.~P.}
\newblock \bibinfo{journal}{\bibinfo{title}{{Evidence for the conjecture that
  sampling generalized cat states with linear optics is hard}}}.
\newblock {\emph{\JournalTitle{Phys. Rev. A}}} \textbf{\bibinfo{volume}{91}},
  \bibinfo{pages}{012342} (\bibinfo{year}{2015}).

\bibitem{Motes2014}
\bibinfo{author}{Motes, K.~R.}, \bibinfo{author}{Gilchrist, A.},
  \bibinfo{author}{Dowling, J.~P.} \& \bibinfo{author}{Rohde, P.~P.}
\newblock \bibinfo{journal}{\bibinfo{title}{{Scalable Boson Sampling with
  Time-Bin Encoding Using a Loop-Based Architecture}}}.
\newblock {\emph{\JournalTitle{Phys. Rev. Lett.}}}
  \textbf{\bibinfo{volume}{113}}, \bibinfo{pages}{120501}
  (\bibinfo{year}{2014}).

\bibitem{Pant2016}
\bibinfo{author}{Pant, M.} \& \bibinfo{author}{Englund, D.}
\newblock \bibinfo{journal}{\bibinfo{title}{{High-dimensional unitary
  transformations and boson sampling on temporal modes using dispersive
  optics}}}.
\newblock {\emph{\JournalTitle{Phys. Rev. A}}} \textbf{\bibinfo{volume}{93}},
  \bibinfo{pages}{043803} (\bibinfo{year}{2016}).

\bibitem{Shen2014}
\bibinfo{author}{Shen, C.}, \bibinfo{author}{Zhang, Z.} \&
  \bibinfo{author}{Duan, L.~M.}
\newblock \bibinfo{journal}{\bibinfo{title}{{Scalable implementation of boson
  sampling with trapped ions}}}.
\newblock {\emph{\JournalTitle{Phys. Rev. Lett.}}}
  \textbf{\bibinfo{volume}{112}}, \bibinfo{pages}{050504}
  (\bibinfo{year}{2014}).

\bibitem{Peropadre2015}
\bibinfo{author}{Peropadre, B.}, \bibinfo{author}{Guerreschi, G.~G.},
  \bibinfo{author}{Huh, J.} \& \bibinfo{author}{Aspuru-Guzik, A.}
\newblock \bibinfo{journal}{\bibinfo{title}{{Proposal for Microwave Boson
  Sampling}}}.
\newblock {\emph{\JournalTitle{Phys. Rev. Lett.}}}
  \textbf{\bibinfo{volume}{117}}, \bibinfo{pages}{140505}
  (\bibinfo{year}{2016}).

\bibitem{rahimi2015}
\bibinfo{author}{Rahimi-Keshari, S.}, \bibinfo{author}{Lund, A.~P.} \&
  \bibinfo{author}{Ralph, T.~C.}
\newblock \bibinfo{journal}{\bibinfo{title}{{What can quantum optics say about
  complexity theory?}}}
\newblock {\emph{\JournalTitle{Phys. Rev. Lett.}}}
  \textbf{\bibinfo{volume}{114}}, \bibinfo{pages}{060501}
  (\bibinfo{year}{2015}).

\bibitem{huh2015}
\bibinfo{author}{Huh, J.}, \bibinfo{author}{Guerreschi, G.~G.},
  \bibinfo{author}{Peropadre, B.}, \bibinfo{author}{McClean, J.~R.} \&
  \bibinfo{author}{Aspuru-Guzik, A.}
\newblock \bibinfo{journal}{\bibinfo{title}{{Boson sampling for molecular
  vibronic spectra}}}.
\newblock {\emph{\JournalTitle{Nature Photon.}}} \textbf{\bibinfo{volume}{9}},
  \bibinfo{pages}{615} (\bibinfo{year}{2015}).

\bibitem{PhysRevA.90.063836}
\bibinfo{author}{Tamma, V.} \& \bibinfo{author}{Laibacher, S.}
\newblock \bibinfo{journal}{\bibinfo{title}{Multiboson correlation
  interferometry with multimode thermal sources}}.
\newblock {\emph{\JournalTitle{Phys. Rev. A}}} \textbf{\bibinfo{volume}{90}},
  \bibinfo{pages}{063836} (\bibinfo{year}{2014}).

\bibitem{Shen2017}
\bibinfo{author}{Shen, Y.}, \bibinfo{author}{Huh, J.}, \bibinfo{author}{Lu,
  Y.}, \bibinfo{author}{Zhang, J.}, \bibinfo{author}{Zhang, K.},
  \bibinfo{author}{Zhang, S.} \& \bibinfo{author}{Kim, K.}
\newblock \bibinfo{journal}{\bibinfo{title}{{Quantum simulation of molecular
  spectroscopy in trapped-ion device}}}.
\newblock {\emph{\JournalTitle{arXiv:1702.04859}}}  (\bibinfo{year}{2017}).

\bibitem{Reck94}
\bibinfo{author}{Reck, M.}, \bibinfo{author}{Zeilinger, A.},
  \bibinfo{author}{Bernstein, H.~J.} \& \bibinfo{author}{Bertani, P.}
\newblock \bibinfo{journal}{\bibinfo{title}{Experimental realization of any
  discrete unitary operator}}.
\newblock {\emph{\JournalTitle{Phys. Rev. Lett.}}}
  \textbf{\bibinfo{volume}{73}}, \bibinfo{pages}{58--61}
  (\bibinfo{year}{1994}).

\bibitem{Braunstein2005}
\bibinfo{author}{Braunstein, S.~L.}
\newblock \bibinfo{journal}{\bibinfo{title}{{Squeezing as an irreducible
  resource}}}.
\newblock {\emph{\JournalTitle{Phys. Rev. A}}} \textbf{\bibinfo{volume}{71}},
  \bibinfo{pages}{055801} (\bibinfo{year}{2005}).

\bibitem{Cariolaro2016}
\bibinfo{author}{Cariolaro, G.} \& \bibinfo{author}{Pierobon, G.}
\newblock \bibinfo{journal}{\bibinfo{title}{{Reexamination of Bloch-Messiah
  reduction}}}.
\newblock {\emph{\JournalTitle{Phys. Rev. A}}} \textbf{\bibinfo{volume}{93}},
  \bibinfo{pages}{062115} (\bibinfo{year}{2016}).

\bibitem{Weedbrook2012}
\bibinfo{author}{Weedbrook, C.}, \bibinfo{author}{Pirandola, S.},
  \bibinfo{author}{Garc{\'{i}}a-Patr{\'{o}}n, R.}, \bibinfo{author}{Cerf,
  N.~J.}, \bibinfo{author}{Ralph, T.~C.}, \bibinfo{author}{Shapiro, J.~H.} \&
  \bibinfo{author}{Lloyd, S.}
\newblock \bibinfo{journal}{\bibinfo{title}{{Gaussian quantum information}}}.
\newblock {\emph{\JournalTitle{Rev. Mod. Phys.}}}
  \textbf{\bibinfo{volume}{84}}, \bibinfo{pages}{621--669}
  (\bibinfo{year}{2012}).

\bibitem{Adesso2014}
\bibinfo{author}{Adesso, G.}, \bibinfo{author}{Ragy, S.} \&
  \bibinfo{author}{Lee, A.~R.}
\newblock \bibinfo{journal}{\bibinfo{title}{{Continuous Variable Quantum
  Information: Gaussian States and Beyond}}}.
\newblock {\emph{\JournalTitle{Open Systems {\&} Information Dynamics}}}
  \textbf{\bibinfo{volume}{21}}, \bibinfo{pages}{1440001}
  (\bibinfo{year}{2014}).

\bibitem{Ma1990}
\bibinfo{author}{Ma, X.} \& \bibinfo{author}{Rhodes, W.}
\newblock \bibinfo{journal}{\bibinfo{title}{Multimode squeeze operators and
  squeezed states}}.
\newblock {\emph{\JournalTitle{Phys. Rev. A}}} \textbf{\bibinfo{volume}{41}},
  \bibinfo{pages}{4625} (\bibinfo{year}{1990}).

\bibitem{duschinsky:1937}
\bibinfo{author}{Duschinsky, F.}
\newblock \bibinfo{journal}{\bibinfo{title}{{Z}ur {D}eutung der
  {E}lektronenspektren mehratomiger {M}olek\"ule}}.
\newblock {\emph{\JournalTitle{Acta Physicochim. URSS}}}
  \textbf{\bibinfo{volume}{7}}, \bibinfo{pages}{551--566}
  (\bibinfo{year}{1937}).

\bibitem{Dierksen2005}
\bibinfo{author}{Dierksen, M.} \& \bibinfo{author}{Grimme, S.}
\newblock \bibinfo{journal}{\bibinfo{title}{{An efficient approach for the
  calculation of Franck-Condon integrals of large molecules}}}.
\newblock {\emph{\JournalTitle{Journal of Chemical Physics}}}
  \textbf{\bibinfo{volume}{122}} (\bibinfo{year}{2005}).

\bibitem{Mann1989}
\bibinfo{author}{Mann, A.}, \bibinfo{author}{Revzen, M.},
  \bibinfo{author}{Nakamura, K.}, \bibinfo{author}{Umezawa, H.} \&
  \bibinfo{author}{Yamanaka, Y.}
\newblock \bibinfo{journal}{\bibinfo{title}{{Coherent and thermal coherent
  state}}}.
\newblock {\emph{\JournalTitle{J. Math. Phys.}}} \textbf{\bibinfo{volume}{30}},
  \bibinfo{pages}{2883} (\bibinfo{year}{1989}).

\bibitem{berger:1998}
\bibinfo{author}{Berger, R.}, \bibinfo{author}{Fischer, C.} \&
  \bibinfo{author}{Klessinger, M.}
\newblock \bibinfo{journal}{\bibinfo{title}{Calculation of the vibronic fine
  structure in electronic spectra at higher temperatures. 1. benzene and
  pyrazine}}.
\newblock {\emph{\JournalTitle{J. Phys. Chem.}}}
  \textbf{\bibinfo{volume}{102}}, \bibinfo{pages}{7157} (\bibinfo{year}{1998}).

\bibitem{Huh2011a}
\bibinfo{author}{Huh, J.}
\newblock \emph{\bibinfo{title}{{Unified description of vibronic transitions
  with coherent states}}}.
\newblock Ph.D. thesis (\bibinfo{year}{2011}).
\newblock
  \eprint{http://publikationen.stub.uni-frankfurt.de/frontdoor/index/index/docId/21033}.

\bibitem{doktorov:1977}
\bibinfo{author}{Doktorov, E.~V.}, \bibinfo{author}{Malkin, I.~A.} \&
  \bibinfo{author}{Man'ko, V.~I.}
\newblock \bibinfo{journal}{\bibinfo{title}{Dynamical symmetry of vibronic
  transitions in polyatomic molecules and the {F}ranck-{C}ondon principle}}.
\newblock {\emph{\JournalTitle{J. Mol. Spectrosc.}}}
  \textbf{\bibinfo{volume}{64}}, \bibinfo{pages}{302--326}
  (\bibinfo{year}{1977}).

\bibitem{jankowiak:2007}
\bibinfo{author}{Jankowiak, H.-C.}, \bibinfo{author}{Stuber, J.~L.} \&
  \bibinfo{author}{Berger, R.}
\newblock \bibinfo{journal}{\bibinfo{title}{Vibronic transitions in large
  molecular systems: Rigorous prescreening conditions for {F}ranck-{C}ondon
  factors}}.
\newblock {\emph{\JournalTitle{J. Chem. Phys.}}}
  \textbf{\bibinfo{volume}{127}}, \bibinfo{pages}{234101}
  (\bibinfo{year}{2007}).

\bibitem{Lee2009}
\bibinfo{author}{Lee, C.-L.}, \bibinfo{author}{Yang, S.-H.},
  \bibinfo{author}{Kuo, S.-Y.} \& \bibinfo{author}{Chang, J.-L.}
\newblock \bibinfo{journal}{\bibinfo{title}{{A general formula of
  two-dimensional Franck–Condon integral and the photoelectron spectroscopy
  of sulfur dioxide}}}.
\newblock {\emph{\JournalTitle{J. Mol. Spectros.}}}
  \textbf{\bibinfo{volume}{256}}, \bibinfo{pages}{279--286}
  (\bibinfo{year}{2009}).

\bibitem{nimlos1986}
\bibinfo{author}{Nimlos, M.} \& \bibinfo{author}{Ellison, G.}
\newblock \bibinfo{journal}{\bibinfo{title}{{Photoelectron spectroscopy of
  sulfur-containing anions (SO2-, S3-, and S2O-)}}}.
\newblock {\emph{\JournalTitle{J. Phys. Chem.}}} \textbf{\bibinfo{volume}{90}},
  \bibinfo{pages}{2574--2580} (\bibinfo{year}{1986}).

\bibitem{glauber:1963}
\bibinfo{author}{Glauber, R.~J.}
\newblock \bibinfo{journal}{\bibinfo{title}{Coherent and incoherent states of
  the radiation field}}.
\newblock {\emph{\JournalTitle{Phys. Rev.}}} \textbf{\bibinfo{volume}{131}},
  \bibinfo{pages}{2766--2788} (\bibinfo{year}{1963}).

\bibitem{sudarshan1963}
\bibinfo{author}{Sudarshan, E. C.~G.}
\newblock \bibinfo{journal}{\bibinfo{title}{Equivalence of semiclassical and
  quantum mechanical descriptions of statistical light beams}}.
\newblock {\emph{\JournalTitle{Phys. Rev. Lett.}}}
  \textbf{\bibinfo{volume}{10}}, \bibinfo{pages}{277} (\bibinfo{year}{1963}).

\end{thebibliography}

\section*{Acknowledgements}

J.H. acknowledges supports by Basic Science Research Program through the National Research Foundation of Korea (NRF) funded by the Ministry of Education, Science and Technology (NRF-2015R1A6A3A04059773), the ICT R\&D program of MSIP/IITP [1711028311] and Mueunjae Institute for Chemistry (MIC) postdoctoral fellowship. M.-H.Y. acknowledges support by the National Natural Science Foundation of China under Grants No. 11405093.

\section*{Author contributions statement}

J.H. conceived and designed the experiments, worked on the theory, and wrote the paper. 
M.-H.Y. worked on the theory and wrote the paper. 

\section*{Correspondence} 
Correspondence and requests for materials should be addressed to J.H. and M.-H.Y.

\section*{Competing financial interests}
The authors declare no competing financial interests.

\end{document}